\documentclass{article}

\RequirePackage{amsthm,amsmath,amsfonts,amssymb}
\RequirePackage[authoryear]{natbib}
\RequirePackage[colorlinks,citecolor=blue,urlcolor=blue,linkcolor = blue]{hyperref}
\RequirePackage{graphicx}
\RequirePackage{icomma}
\RequirePackage{multirow}
\usepackage{array}
\usepackage{rotating}
\usepackage[section]{placeins}
\usepackage{bm}
\usepackage[a4paper, margin = 1in]{geometry}

\RequirePackage{amsthm,amsmath,amsfonts,amssymb}
\RequirePackage[authoryear]{natbib}
\RequirePackage[colorlinks,citecolor=blue,urlcolor=blue]{hyperref}
\RequirePackage{graphicx}
\RequirePackage{bm}
\RequirePackage{multirow}
\RequirePackage{rotating}

\DeclareMathOperator*{\argmax}{argmax}
\DeclareMathOperator*{\argmin}{argmin}

\begin{document}


\newcommand{\E}{\text{E}}
\newcommand{\MVN}{\text{MVN}}
\newcommand{\N}{\text{N}}
\newcommand{\Unif}{\text{Unif}}
\newcommand{\Bern}{\text{Bern}}
\newcommand{\ivar}[1]{\text{I}\left\{#1\right\}}

\newcommand{\Yij}{Y_{ij}}
\newcommand{\yij}{y_{ij}}
\newcommand{\Xij}{\bm{X_{ij}}}
\newcommand{\Xijk}[1]{x_{ij#1}}
\newcommand{\Aij}{{A_{ij}}}
\newcommand{\aij}{{a_{ij}}}
\newcommand{\ti}[1]{t_{i#1}}
\newcommand{\xij}{\bm{x_{ij}}}
\newcommand{\xijo}{x_{ijk_1}}
\newcommand{\xijoo}{x_{ijk_2}}

\newcommand{\xp}{x_{1}, \dots, x_Q}
\newcommand{\xx}{\bm{x}}

\newcommand{\Mij}{M_{ij}}
\newcommand{\mij}{m_{ij}}

\newcommand{\ddeltaik}{\bm{\delta}_{ik}}
\newcommand{\ddeltai}{\bm{\delta}_{i}}
\newcommand{\bbetai}{\bm{\beta}_i}
\newcommand{\aalphai}{\bm{\alpha}_i}
\newcommand{\tthetai}{\bm{\theta}_i}

\newcommand{\ppsi}{\bm{\psi}}
\newcommand{\SSigma}{\bm{\Sigma}_i}

\newcommand{\hddeltai}{\hat{\bm{\delta}}_{i}}
\newcommand{\hbbetai}{\hat{\bm{\beta}}_{i}}

\newcommand{\pit}{\pi^{t}}
\newcommand{\pim}{\pi^{m}}

\newcommand{\xijb}{\bm{x_{ij}^{\beta}}}
\newcommand{\xijd}{\bm{x_{ij}^{\delta}}}
\newcommand{\xijm}{\bm{x_{ij}^{\theta}}}
\newcommand{\xijt}{\bm{x_{ij}^{\alpha}}}

\newcommand{\Ui}{\bm{U}_i}
\newcommand{\ident}[1]{\bm{I}_{(#1)}}
\newcommand{\Vi}{\bm{V}_i}

\newcommand{\Si}{\hat{\bm{\Sigma}}(\hddeltai)}

\newcommand{\alphal}{\aalphai^{(l)}}
\newcommand{\thetal}{\tthetai^{(l)}}
\newcommand{\deltalk}{\ddeltaik^{(l)}}
\newcommand{\deltal}{\ddeltai^{(l)}}
\newcommand{\betal}{\bbetai^{(l)}}
\newcommand{\omglj}{\omega_{lj}}
\newcommand{\sumj}{\sum_{j = 1}^{\Ni}}
\newcommand{\sumk}{\sum_{k = 2}^{G_i}}
\newcommand{\Ni}{N_{i}}
\newcommand{\Gi}{G_{i}}
\newcommand{\oomgl}{\bm{\omega}^{(l)}}

\title{Doubly-Robust Bayesian Estimation of Optimal Individualized Treatment Rules Using Network Meta-Analysis}

\author{Augustine Wigle \and Erica E. M. Moodie}

\maketitle

\begin{abstract}

An optimal individualized treatment rule (ITR) is a function that takes a patient's characteristics, such as demographics, biomarkers, and treatment history, and outputs a treatment that is expected to give the best outcome for that patient. Major Depressive Disorder (MDD) is a common and disabling mental health condition for which an optimal ITR is of interest. Unfortunately, the power to detect treatment-covariate interactions in individual studies of MDD treatments is low. Additionally, all treatments of interest are not compared head-to-head in a single study. Network meta-analysis (NMA) is a method of synthesizing data from multiple studies to estimate the relative effects of a set of treatments. Recently, two-stage ITR NMA was proposed as a method to estimate ITRs that has the potential to improve power and simultaneously consider all relevant treatment options. In the first stage, study-specific ITRs are estimated, and in the second stage, they are pooled using a Bayesian NMA model. The existing approach is vulnerable to model misspecification and fails to address missing outcomes, which occur in the MDD data. We overcome these challenges by proposing Bayesian Bootstrap dynamic Weighted Ordinary Least Squares (BBdWOLS), a doubly-robust approach to ITR estimation that accounts for missing at random outcomes and naturally quantifies the uncertainty in estimation. We also propose an improvement to the NMA model that incorporates the full variance-covariance matrix of study-specific estimates. In a simulation study, we show that our fully Bayesian ITR NMA method is more robust and efficient than the existing approach. We apply our method to the motivating dataset consisting of three studies of pharmacological treatments for MDD, and explore how ITR NMA results can support personalized decision making in this context.
\end{abstract}

\section{Introduction} 

An optimal individualized treatment rule (ITR) is a function that outputs the treatment that is expected to give the best outcome given a patient's characteristics out of a set of candidate treatments. ITRs are useful when treatment effects are heterogeneous, so that the ITR outputs different treatments depending on the value of effect-modifying variables. Major Depressive Disorder (MDD) is common and a leading cause of disability \citep{perlman_systematic_2019}. Symptoms, underlying causes, and responses to the many available treatments for MDD can vary widely among individuals, motivating researchers to seek an optimal ITR for MDD that can map to a broad set of treatments \citep{kessler_using_2017,perlman_systematic_2019}. 

ITR estimation using data from a single study comes with challenges. First, the power to detect all important treatment-covariate interactions is typically low in any given study \citep{greenland_tests_1983, brookes_subgroup_2004}. Second, an ITR estimated from a single study only compares the performance of the treatments included in the study, and it is rare for a single study to encompass all relevant candidate treatments. Pooling data from multiple studies can help overcome both of these challenges but presents concerns about data privacy if the method requires sharing sensitive individual participant data (IPD). Methods which allow multi-study data to be pooled without requiring the release of IPD are thus particularly attractive. 

Network meta-analysis (NMA) is a method used to pool data from multiple randomized studies that compare different but overlapping sets of treatments \citep{lumley_network_2002}. \citet{shen_two-stage_2025} recently proposed a two-stage NMA approach to estimating ITRs using multi-study data. This approach estimates  study-level ITRs in the first stage, then pools the results using a Bayesian NMA model in the second stage. The optimal ITR estimated in the second stage maps to the full set of treatments in the network. The two-stage approach avoids the need to release IPD because only the study-level ITR estimates need to be released for use in the NMA model. 

The existing approach to ITR NMA is sensitive to misspecification of the models used to estimate ITRs in stage one. Missing outcomes are also not accommodated, although they are common in practice, including in studies of MDD. In stage two, the method does not leverage the full covariance matrix of study-level estimates. It has been acknowledged that ignoring correlation in the data in meta-analysis models can change relative effect estimates and their precision \citep{riley_multivariate_2009}. The first aim of this work is to provide a fully Bayesian doubly-robust approach to multi-site ITR NMA that accommodates missing outcome data and utilises the full data covariance matrix. The second aim is to use our proposed approach to estimate an optimal ITR for MDD.

The rest of this article is structured as follows: In Section \ref{sec:data}, we describe a motivating dataset consisting of three studies that investigated treatments for MDD. In Section \ref{sec:notation}, we introduce the necessary notation and background on ITR estimation. We extend dynamic weighted ordinary least squares (dWOLS), a standard frequentist ITR estimation approach, to account for missing outcomes in Section \ref{sec:dwols} and we propose Bayesian Bootstrap dWOLS (BBdWOLS), a robust Bayesian approach to study-level ITR estimation that incorporates missing outcome data and provides natural accounting of the uncertainty in all aspects of the estimation procedure in Section \ref{sec:BBdWOLS-miss}. We then describe a Bayesian ITR NMA model for synthesizing study-level ITR estimates in Section \ref{sec:nmamodel}. We perform a simulation study investigating the properties of using BBdWOLS and the proposed ITR NMA model in Section \ref{sec:sim}. In Section \ref{sec:dataresults}, we use the proposed methods to estimate an optimal ITR for MDD using the motivating dataset. We conclude with a discussion of our results and areas for development in Section \ref{sec:disc}.

\section{Motivating Data}\label{sec:data}


We use data from three multi-stage clinical trials that compared a wide set of treatments aimed at reducing depressive symptoms for patients with MDD. The studies are Establishing Moderators and Biosignatures of Antidepressant Response for Clinical Care (EMBARC) \citep{trivedi_establishing_2016}, Research Evaluating the Value of Augmenting Medication with Psychotherapy (REVAMP) \citep{trivedi_revamp_2008}, and Sequenced Treatment Alternatives to Relieve Depression (STAR*D) \citep{rush_sequenced_2004}. The outcome of interest is depression severity measured using the 17-item Hamilton Rating Score for Depression (HRSD-17). The studies investigated pharmacological, psychological, and combination treatments, but the number of unique psychological and combination treatments is very high, leading to a very sparse network. Thus, we focus our investigation on active pharmacological treatments. The studies considered subsets of the following pharmacological treatments: Sertraline (SER), Bupropion (BUP), Escitalopram (ESCIT), Venlafaxine (VEN), Citalopram and Bupropion together (CIT + BUP), and Citalopram and Buspirone together (CIT + BUS).

EMBARC is a two-stage trial where participants are randomized to either placebo or SER in the first stage. In the second stage, participants who did not respond to treatment in the first stage switch treatments, with those who were initially randomized to placebo switching to SER and those initially randomized to SER switching to BUP. Since data from stage one will not provide information on the relative performance of active treatments, we use data from the second stage for the participants who did not respond in stage one. The data from EMBARC thus provides information on treatments SER and BUP.

REVAMP is a two-stage trial. In the first stage, participants are assigned to one of SER, ESCIT, BUP, or VEN based on the treatments they have tried in the past two years to enable broad inclusion criteria and avoid assigning a participant to a treatment which is not effective for them. In the second stage, participants are randomized to receive different psychotherapies in combination with medication or to continue with medication only. Since our interest is in comparing pharmacological treatments for MDD, we use data from stage one. Only 11 participants received VEN in stage one, leading to unstable relative effect estimates. Thus, REVAMP ultimately contributes information on SER, ESCIT, and BUP. 

STAR*D is a Sequential Multiple Assignment Randomized Trial well-suited to ITR estimation. In the first stage, all participants received CIT. Participants who do not respond satisfactorily to CIT moved on to the second stage, where they were randomized to a treatment corresponding to switching to a new medication (one of SER, BUP, or VEN),  augmenting CIT with a medication (add one of BUS or BUP), switching to psychotherapy, or augmenting with psychotherapy. Participants indicated which of medication switch, medication augmentation, psychotherapy switch and psychotherapy augmentation were acceptable to them and they were randomized to a new treatment regime from among the acceptable options. We use stage two data from participants who deemed one or both of medication switch or medication augmentation as acceptable and who were not randomized to a treatment involving psychotherapy. STAR*D provides information on SER, BUP, VEN, CIT + BUP, and CIT + BUS. 

\begin{figure}[tb]
    \centering
    \includegraphics[width=0.6\linewidth]{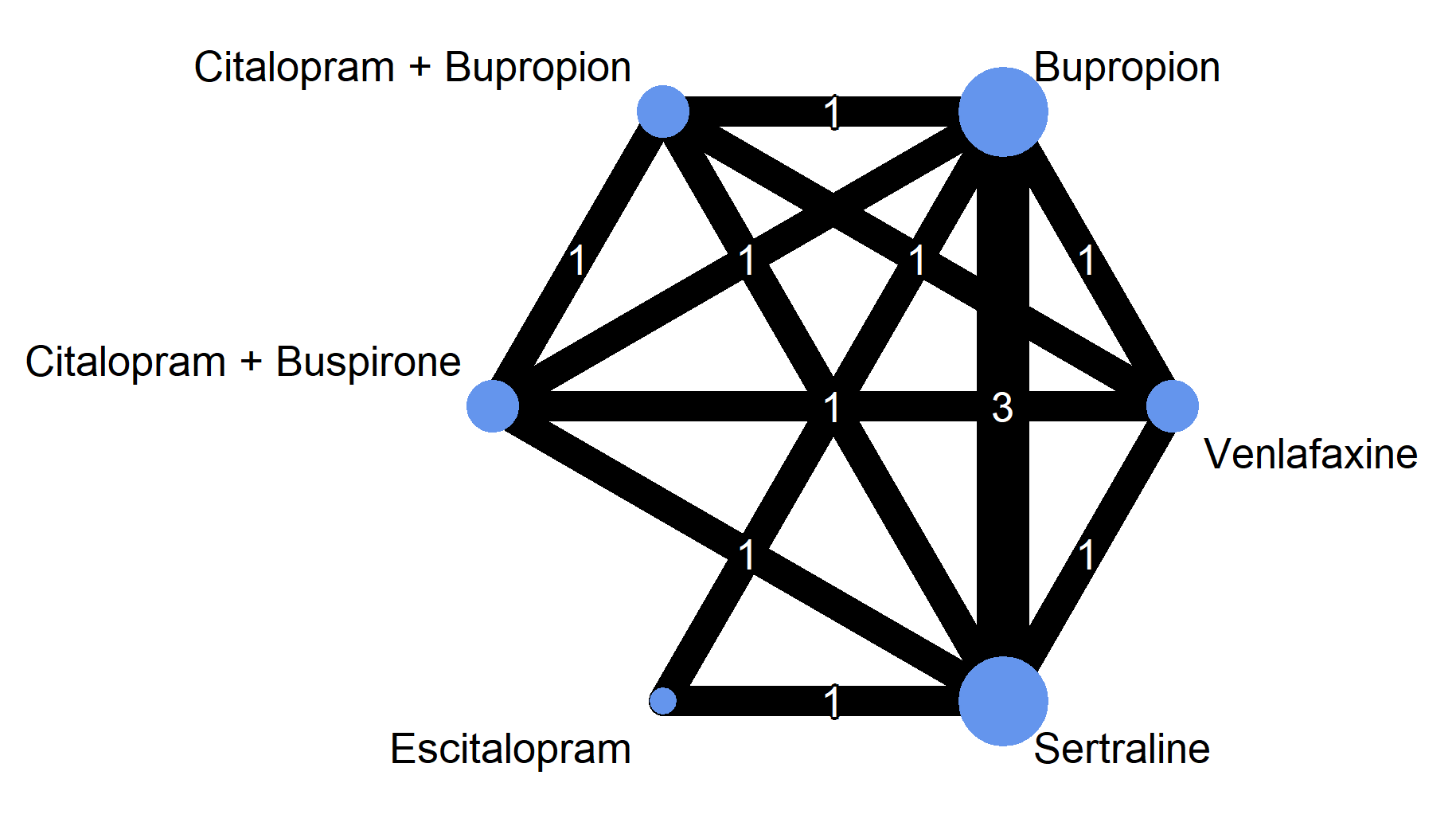}
    \caption{Network plot of the MDD studies (EMBARC, REVAMP, and STAR*D). Each node represents a treatment and each edge represents head-to-head evidence comparing two treatments. The size of each node is proportional to the total number of individuals in the network who received that treatment. The number of studies making each direct comparison is written on each edge.}
    \label{fig:datanetwork}
\end{figure}

Together, the data create the network of six treatments shown in Figure \ref{fig:datanetwork}, meaning we can pool the studies to create an ITR that maps to these six treatments. In the EMBARC study, participants had to have an age of MDD onset as 30 years or less. We thus excluded participants from REVAMP and STAR*D  who had an age of onset greater than 30 years to improve overlap of the populations. The meta-population of the data, that is, the union of the study populations, represents MDD patients who may have failed on another treatment in the current episode and who had an onset of MDD on or before age 30. EMBARC contributes data from 126 individuals, REVAMP contributes data from 429 individuals and STAR*D contributes data from 871 individuals. In addition to 22-33\% of individuals in each study missing their HRSD-17 at the end of the stage, some prognostic or effect-modifying covariates are also missing for some patients in each study. The percentage of missing values for the response and covariates in each study are listed in Table \ref{tab:missingperc}. Of note are the number of MDD episodes, which is missing for 11.2\% of participants in REVAMP and 19.3\% STAR*D, and the baseline HRSD-17 score, which is missing for 12.6\% of participants in REVAMP and 1.3\% in STAR*D. 

\begin{table}[tb]
    \centering
    \caption{Percentage of participants in each study with missing values for given covariates of interest.}
    \label{tab:missingperc}
    \begin{tabular}{l|rrr}
        Variable & \% Missing EMBARC & \% Missing REVAMP & \% Missing STAR*D \\ \hline
        Response & 22.2 & 33.1 & 27.6 \\
        Race & 0.0 & 0.0 & 8.7 \\
        Employment status & 2.4 & 0.5 & 0.2 \\
        Years of formal education & 0.0 & 0.5 & 0.3 \\
        Marital status & 0.0 & 0.2 & 0.0 \\
        Number MDD episodes &  1.6 & 11.2 & 19.3 \\
        Baseline HRSD-17 & 0.0 & 12.6 & 1.3 \\
        Chronicity of current episode & 0.8 & 0.5 & 0.0 \\
    \end{tabular}
\end{table}

\section{Methods} 

\subsection{Notation and Background}\label{sec:notation}

We consider the setting where the outcome of interest for individual $j$ in study $i$, $\Yij$, is a continuous random variable, where a larger value indicates a more preferred outcome. We also have available covariates $\Xij$, which includes prognostic and effect-modifying variables, and the treatment assigned to individual $j$, $\Aij$. We use uppercase and lowercase to denote random variables and their realizations respectively, and bold font to denote vectors.

In the following, we make three assumptions common in the causal literature: i) the stable unit treatment value assumption, that is, that a participant's outcome is not influenced by the treatment assignment of other participants, ii) there are no unmeasured confounding variables, and iii) positivity, that is, every possible covariate combination could have in principle received any of the treatments within each study \citep{tsiatis_dynamic_2019, cole_constructing_2008, shen_two-stage_2025}. Since our focus in this work is on pharmacotherapy rather than, say, group psychotherapy, i) is likely to hold. Further, in randomized trial data, ii) and iii) hold trivially, though we return to point ii) in the discussion.

To estimate an optimal ITR from a single study, we model the average response of individual $j$ in study $i$ with sample size $N_i$ as
\begin{equation}
    \E(\Yij \mid \Xij = \xij, \Aij = \aij; \bbetai, \ddeltaik) = \bbetai^T\xijb + \sumk \ivar{\aij = \ti{k}} {\ddeltaik}^T \xijd, \label{eq:regmod}
\end{equation}
where $\xijb$ and $\xijd$ are (sub)vectors of $\xij$ containing appropriate covariates and leading ones, $\ti{k}$ represents the treatment in arm $k$ of study $i$, $\Gi$ is the number of treatments available in study $i$, and $\bbetai$ and $\ddeltaik$ are parameter vectors. The term $\bbetai^T\xijb$ in Equation \eqref{eq:regmod} describes the expected response for individual $j$ to the treatment in arm 1 of study $i$, and we refer to this part of the model as the \textit{reference} model. The second term of Equation \eqref{eq:regmod} is called the \textit{blip} model, and describes how the relative effects are expected to change for different treatments, that is, the blip models treatment contrasts. The vector $\xijb$ includes all prognostic and effect-modifying variables and $\xijd$ is a subvector of $\xijb$ that contains all effect-modifiers. 

The elements of vector $\ddeltaik$ can be written as
\begin{equation*}
    \ddeltaik = (\delta_{i, \ti{k}\ti{1}, 0}, \delta_{i, \ti{k}\ti{1},1}, \dots, \delta_{i, \ti{k}\ti{1},Q})', 
\end{equation*}
where $\delta_{i, \ti{k}\ti{1},0}$ is the expected change in the response for a patient that receives treatment $\ti{k}$ compared to treatment $\ti{1}$ in study $i$ when all covariates are equal to zero, and $\delta_{i, \ti{k}\ti{1},q}, q \geq 1$ is the interaction of covariate $q$ with treatment $\ti{k}$ compared to treatment $\ti{1}$ in study $i$. By definition, $\bm{\delta}_{i1}$ is a vector of zeroes and need not be estimated. Assuming correct specification of the blip models in Equation (1), the true optimal ITR is given by
\begin{equation*}
    a^{opt} = \argmax_{k \in 1, \dots, \Gi}  {\ddeltaik}^T \xijd.
\end{equation*}

Let $\ddeltai = (\bm{\delta}_{i2}', \dots, \bm{\delta}_{i\Gi}')'$ be a vector of length $(Q+1)\times(\Gi-1)$ that stores the true blip parameters in study $i$. Identifying the optimal ITR amounts to estimating $\ddeltai$, which is possible through a variety of regression-based approaches, such as Q-learning or dWOLS \citep{c_j_c_h_watkins_learning_1989, sutton_reinforcement_1998, wallace_doubly-robust_2015, shen_two-stage_2025}. Q-learning estimates blip parameters by fitting an ordinary least squares model of the form \eqref{eq:regmod} to study data. This method provides consistent estimators of $\ddeltai$ when the causal assumptions outlined earlier in this section are satisfied and both the blip model and reference model are correctly specified. Note that equation \eqref{eq:regmod} is linear in parameters, but non-linear transformations of covariates can be included in $\xijb$ and $\xijd$ (e.g., polynomials or splines) to ensure flexible specification of the reference and blip models. 

A more robust alternative to Q-learning is dWOLS. It involves fitting a weighted ordinary least squares regression based on the model in Equation \eqref{eq:regmod} where the weights satisfy a balancing condition that removes bias due to confounding between covariates and treatment assignment \citep{wallace_doubly-robust_2015, schulz_doubly_2021}. In particular, for categorical treatments, weights of the form
\begin{equation*}
    \frac{1}{\Pr(\Aij = \aij \mid \xijt; \aalphai)} = \frac{1}{\pit_{ij} (\aalphai)}
\end{equation*}
satisfy the balancing condition, where $\xijt$ is a (sub)vector of $\xij$ and $\aalphai$ is a vector of parameters for the treatment assignment model \citep{schulz_doubly_2021}. In practice, the parameters of the treatment assignment model are estimated and $\hat{\aalphai}$ is plugged in to obtain weights 
\begin{equation} \label{eq:dwols-wts}
     \frac{1}{\pit_{ij} (\hat{\aalphai})}.
\end{equation}

With the weights given in \eqref{eq:dwols-wts}, dWOLS provides consistent estimators of $\ddeltai$ when the blip model is correctly specified, standard causal assumptions outlined earlier in this section are satisfied, and one or both of the following two conditions are met: 1. The reference model is correctly specified, or 2. 
The treatment assignment model is correctly specified with respect to confounding variables. The latter condition is trivially satisfied in the context of randomized treatment assignment. This flexibility makes dWOLS doubly-robust, and in particular, more robust than Q-learning.


\subsection{Estimating Study-Level ITRs via dWOLS with Missing Outcomes} \label{sec:dwols}

 Neither Q-learning nor dWOLS account for missing outcomes, which are prevalent in practice, such as in the dataset described in Section \ref{sec:data}. Missing outcomes can lead to biased estimation if not accounted for properly, so we extend dWOLS to the case where outcomes are missing at random (MAR), that is, missingness is related to covariates and treatment assignment, which are all completely observed.

Let $\Mij$ be a random variable that equals $1$ if individual $j$ in study $i$ is missing the outcome and $0$ otherwise. If the outcome is independent of missingness given treatment and covariates, that is, outcomes are MAR, then $\E(\Yij \mid \xij, \aij, \Mij = 0) = \E(\Yij \mid \xij, \aij)$, so a weighted regression of the complete cases of the form \eqref{eq:regmod} will give consistent estimators of the blip parameters in study $i$ as long as the reference model and blip model are correctly specified, in line with standard dWOLS. However, if the reference model is misspecified, the weights in \eqref{eq:dwols-wts} will not adequately adjust for confounding if $\Mij$ depends on $\xij$ and $\aij$. Thus, we propose combined weights,
\begin{equation}\label{eq:mar-wts}
    \left[\pim_{ij}(\hat{\tthetai}) \pit_{ij}(\hat{\aalphai})\right]^{-1},
\end{equation}
where 
\begin{equation}
    \pim_{ij}(\tthetai) = \Pr(\Mij = 1 \mid \xijm, \aij; \tthetai),
\end{equation}
$\xijm$ is a (sub)vector of $\xij$, $\tthetai$ is a parameter vector, and $\hat{\tthetai}$ is an estimate of $\tthetai$.
A derivation of these weights and the balancing condition for MAR outcomes is given in Supplementary Information (SI) \S 1.

Performing a weighted ordinary least squares based on model \eqref{eq:regmod} with the weights given in \eqref{eq:mar-wts} gives consistent estimators of the blip parameters when the blip model is correctly specified, the outcomes are MAR, the treatment assignment and covariates are completely observed, and one or more of the following two conditions are met: 1. The reference model is correctly specified, or 2. The treatment assignment model and the missingness model used to calculate the weights are both correctly specified.

\subsection{A Bayesian Approach to Study-Level ITR Estimation with Missing Outcomes} \label{sec:BBdWOLS-miss}


{The MAR dWOLS approach to estimating a study-specific ITR outlined in the previous subsection does not account for uncertainty in the estimation of the treatment assignment or missingness model parameters. Bootstrapping, where the observed data in study $i$ with sample size $\Ni$ is re-sampled with replacement and the estimation procedure is performed on the new sample repeatedly, is commonly used to account for parameter uncertainty in dWOLS \citep{efron_bootstrap_1979, wallace_doubly-robust_2015, schulz_doubly_2021}. The classic bootstrap can be interpreted as repeatedly sampling weights for each observation from a Multinomial distribution, where each observed sample value has probability $1/\Ni$ of being selected, resulting in integer weights that sum to $\Ni$ in each bootstrap iteration \citep{xu_applications_2020}. The Bayesian Bootstrap (BB) is a Bayesian analogue to the classic bootstrap \citep{rubin_bayesian_1981, xu_applications_2020}.  In the BB, weights are typically sampled from a Dirichlet(1,...,1) distribution, providing continuous-valued weights, although sampling weights from other distributions is possible \citep{rubin_bayesian_1981,xu_applications_2020, greifer_fwb_2025}. The BB with Dirichlet weights is thus a ``smoothed'' version of the classic bootstrap, and rather than providing an approximation to the sampling distribution of a statistic, provides samples from a posterior distribution of a parameter \citep{rubin_bayesian_1981, xu_applications_2020}. We will use the BB rather than the classic bootstrap in estimating study-level ITRS to enable a cohesive, fully Bayesian approach to ITR NMA.

The algorithm for obtaining the posterior distribution of the blip parameters for study $i$ via BBdWOLS with MAR outcomes is as follows: For $l = 1, \dots, L$, do the following:}
\begin{enumerate} 
        \item Draw a Dirichlet(1, ..., 1) random variable, $\oomgl = (\omega_{l1}, \omega_{l2}, ..., \omega_{l\Ni})$.
        \item Compute the value of the posterior of $\tthetai$ in iteration $l$, $\thetal$, which is the value of $\tthetai$ which maximizes the weighted log likelihood of the missingness model with weights given by $\oomgl$.
        \item Compute the value of the posterior of $\aalphai$ in iteration $l$, $\alphal$, which is the value of $\aalphai$ which maximizes the weighted log likelihood of the treatment assignment model with weights given by $\oomgl$.
        \item Compute the value of the posterior of $(\bbetai, \ddeltai)'$ at iteration $l$ by minimizing the weighted loss function:
        \begin{equation*}
            (\betal, \deltal)' = \argmin_{(\bbetai, \ddeltai)} \sumj w_{ijl}(1-\mij)\left(\yij - \left[{\bbetai}^T\xijb + \sumk \ivar{\aij = \ti{k}} {\ddeltaik}^T \xijd\right]\right)^2,
        \end{equation*}
        where the term $(1-\mij)$ effectively drops the observations with missing outcomes and
        \begin{equation*} \label{eq:finalwts}
            w_{ijl} = \frac{\omglj}{\pit_{ij}( \alphal)\pim_{ij}(\thetal)}.
        \end{equation*}
\end{enumerate}
The sample $\{\ddeltai^{(1)}, \dots, \ddeltai^{(L)} \}$ is an independent sample from the posterior distribution of the blip parameters (and therefore the optimal ITR) given the data. The uncertainty in the estimation of $\aalphai$ and $\tthetai$ are propagated into the posterior distribution. 
The posterior mean or median can be used as a point estimate for  $\ddeltai$, $\hddeltai$.
A posterior estimate of the covariance matrix of the estimates, $\Si$, can be estimated by calculating its sample counterpart from the posterior samples. Credible intervals (CrIs) can be calculated from the sample as desired, for example, using quantile-based CrIs.

\subsection{Bayesian NMA Model for ITRs} \label{sec:nmamodel}

After study-specific ITRs have been estimated using any approach that gives consistent estimates of the blip parameters, we can synthesize the results using ITR NMA. Let $I$ represent the number of studies we wish to synthesize and let $G$ represent the number of treatments in the full treatment set spanned by the $I$ studies. We introduce $\psi_{gh,q}$, $g,h= 1, \dots, G$, the expected relative effect of treatment $g$ compared to treatment $h$ in the meta-population when all effect modifiers are zero for $q = 0$ and the expected interaction of covariate $q$ with treatment $g$ compared to $h$ for $q =1, \dots, Q$. By definition, $\psi_{gg,q} = 0$ for all $g,q$. We assume that the meta-population parameters obey the consistency equations, a common assumption in the NMA literature \citep{ades_twenty_2024}; note that this is distinct from the causal consistency assumption. In particular, we assume that
\begin{equation*}
    \psi_{gg',q} = \psi_{gh,q} - \psi_{g'h,q},\quad q = 0, \dots, Q,
\end{equation*}
where $g, g'$, and $h$ are any treatments in the full treatment set.



Due to the NMA consistency assumption, an optimal ITR for the meta-population is fully characterized by $(G-1)\times(Q+1)$ meta-population blip parameters with a common reference, say, the treatment labelled as 1. 

Let
\begin{equation*}
    \ppsi = (\psi_{21,0}, \psi_{21,1}, \dots, \psi_{21,Q}, \psi_{31,0}, \dots, \psi_{31,Q}, \dots, \psi_{G1, 0}, \dots, \psi_{G1,Q})'
\end{equation*}
be the vector of length $(Q+1)\times(G-1)$ that stores the set of true meta-population blip parameters. The  optimal ITR in the meta-population is given by
\begin{equation*}
    \argmax_{g\in 1, \dots, G} \ppsi_g'\bm{x},
\end{equation*}
where $\ppsi_{g}$ is a subvector of $\ppsi$ of length $Q+1$ containing the blip parameters associated with treatment $g$ and $\bm{x}$ is a vector of $Q$ effect modifier values with a leading 1.

Since each study only directly contributes to estimating a subset of $\ppsi$ unless the study investigated all $G$ treatments, we introduce $\Ui$, a matrix of dimension $(\Gi-1)\times (G-1)$ that describes the treatment comparisons reflected by the blip parameters of study $i$. Each row of $\Ui$ represents the comparison of one treatment arm to the study-specific reference arm containing treatment $\ti{1}$. Each column represents one treatment in the full treatment set. In particular, to represent the non-reference treatments, $U_{i,nm} = 1$ if $\ti{n+1} = m + 1$ for $n = 1, \dots, \Gi-2$, and 0 otherwise. If the study-specific reference treatment is not the same as the meta-population reference treatment, then if $\ti{1} = m+1$, then every entry of column $m$ of $\Ui$ is equal to -1. If the study-specific reference treatment is the same as the meta-population reference, that is $\ti{1} = 1$, then there is no need to add a column of $-1$. For example, in a three-arm study that investigated treatments 2, 3, and 4 out of 5 total treatments, where $\ti{1} = 2$, $\Ui$ is given by
\begin{equation*}
    \Ui = \begin{pmatrix}
        -1 & 1 & 0 & 0 \\
        -1 & 0 & 1 & 0 
    \end{pmatrix}.
\end{equation*}
For a two-arm study that investigated treatments 1 and 2, where $\ti{1} = 1$, we have
\begin{equation*}
    \Ui = \begin{pmatrix}
        1 & 0 & 0 & 0
    \end{pmatrix}.
\end{equation*}

The study-specific blip parameters $\ddeltai, i = 1, \dots, I$ are assumed to be exchangeable, with mean vectors defined in terms of $\ppsi$ and the consistency equations. The distribution is given by
\begin{equation*}
    \ddeltai \sim \MVN\left(\Vi \ppsi, \SSigma\right), i = 1, \dots, I,
\end{equation*}
where  
\begin{equation*}
    \Vi = \Ui \otimes \ident{Q+I}
\end{equation*}
is a $\left([Q+1]\times[\Gi-1]\right) \times \left([Q+1] \times [G-1]\right)$ matrix that maps the study-specific parameters in $\ddeltai$ to the meta-population parameters in $\ppsi$ using the consistency equations, $\ident{Q+1}$ is the $(Q+1) \times (Q+1)$ identity matrix, and $\otimes$ is the Kronecker product. The variance matrix of $\ddeltai$ can take different forms, with a common form given by
\begin{equation*}
    \SSigma = \begin{pmatrix}
        \tau^2 & \tau^2/2 & \dots & \tau^2/2 \\
        \tau^2/2 & \tau^2 & \dots & \tau^2/2 \\
        \vdots & \dots & \ddots & \vdots \\
        \tau^2/2 & \dots & \tau^2/2 & \tau^2
    \end{pmatrix}, \quad i = 1, \dots, I,
\end{equation*}
corresponding to the assumption that all treatment comparisons and interaction terms have the same between-study heterogeneity and all studies display the same degree of heterogeneity (the common variance assumption). More complex specifications are possible, but the common variance assumption leads to an important reduction in the number of parameters which can improve model convergence, and was found in a simulation to have good performance even when the assumption was violated \citep{shen_two-stage_2025}. 

The full NMA model can be written as
\begin{equation} \label{eq:fullmod}
\begin{aligned}
    \hddeltai &\sim \MVN(\ddeltai, \Si) \quad i = 1, \dots, I; \\
    \ddeltai &\sim \MVN\left(\Vi \ppsi, \SSigma\right) \quad i = 1, \dots, I;\\
    \ppsi  &\sim f_{\ppsi}, \SSigma \sim f_{\SSigma}, i = 1, \dots, I
\end{aligned}
\end{equation}
where $\Si$ is the estimated variance-covariance matrix of $\hddeltai$ and $f_{\ppsi}$ and $f_{\SSigma}$ are prior distributions reflecting the state of knowledge about $\ppsi$ and $\SSigma$ before synthesizing the data, discussed further in the following subsection. Note that in this formulation, we use the full variance-covariance matrix of the study-specific parameter estimates, while the model proposed by \citet{shen_two-stage_2025} uses a sparse version of $\Si$ where zero covariance is assumed between parameter estimates for different covariate types, for example, $Cov(\hat{\delta}_{i, \ti{k}\ti{1}, q}, \hat{\delta}_{i, \ti{k}\ti{1}, q'}) = 0$ for $q' \neq q$. Our formulation thus leverages additional study-level information which can influence the point estimates and uncertainty in the NMA model. 

The model as written is a random-effects model. To use a common-effects model (one which assumes there is no between-study heterogeneity), replace the second line of Equation \eqref{eq:fullmod} with 
\begin{equation*}
    \ddeltai = \Vi \ppsi, \quad i = 1, \dots, I.
\end{equation*}

\subsubsection{Prior Specification} 

An important component of the Bayesian model is specification of the priors $f_{\ppsi}$ and $f_{\SSigma}$. It is reasonable to expect that the number of studies to be included in an ITR NMA may be quite small since IPD can be difficult to access, meaning that extra care should be taken in prior specification. Common practice in NMA is to assume that treatment relative effects (e.g., $\psi_{gh,0}$) are a priori independent and follow a normal distribution centred at 0 with a large standard deviation, such as 10 or 100 \citep{van_valkenhoef_automating_2012, dias_network_2018}. Centring the prior at zero ensures that any bias from the prior is conservative, that is, biased towards the assumption of no effect, and the large standard deviation keeps the prior relatively uninformative \citep{dias_network_2018}. This logic can also be applied to the interaction parameters, $\psi_{gh,q}, q > 0$. We adopt independent normal priors centred at zero for all elements of $\ppsi$, where the standard deviations are chosen to be large relative to the scale of the data, and external information could be incorporated here, for example, information about the range of a continuous outcome \citep{van_valkenhoef_automating_2012}. 

In random-effects models, the prior distribution for between-trial heterogeneity can have a substantial impact on estimation, especially when there are few repeated observations of treatment contrasts \citep{rover_weakly_2021, dias_network_2018}. \citet{shen_two-stage_2025} provide a brief overview of priors for $\SSigma$. Under the common variance assumption, only a univariate prior for the common between-study heterogeneity standard deviation $\tau$ is needed. We suggest the use of a half-normal prior for $\tau$ where the scale parameter is specified by considering the scale of the data, the implications of different $\tau$ values on $\ddeltaik$, and any other available external information, as described in \citet{rover_weakly_2021}. When there is little information in the data due to study availability, practitioners may have to resort to fitting models with multiple priors for $\tau$ as a sensitivity analysis.

\section{Simulation}\label{sec:sim} 


The double-robustness of BBdWOLS with and without MAR outcomes is established in a detailed simulation study in SI \S 2. We found that the MAR weights proposed in Section \ref{sec:dwols} are robust to model misspecification and lead to closer-to-nominal coverage probability compared to the standard weights in the presence of MAR outcomes. In this section, we conduct a simulation study aimed at showing the advantages of using a doubly-robust approach in the first stage of a two-stage ITR NMA and investigating the impact of 1) the covariance matrix of $\hddeltai$ and 2) the prior for $\tau$ on the NMA results. 

\subsection{Data Generating Mechanisms (DGMs)}

\begin{figure}[tb]
    \centering
    \includegraphics[width=0.5\linewidth]{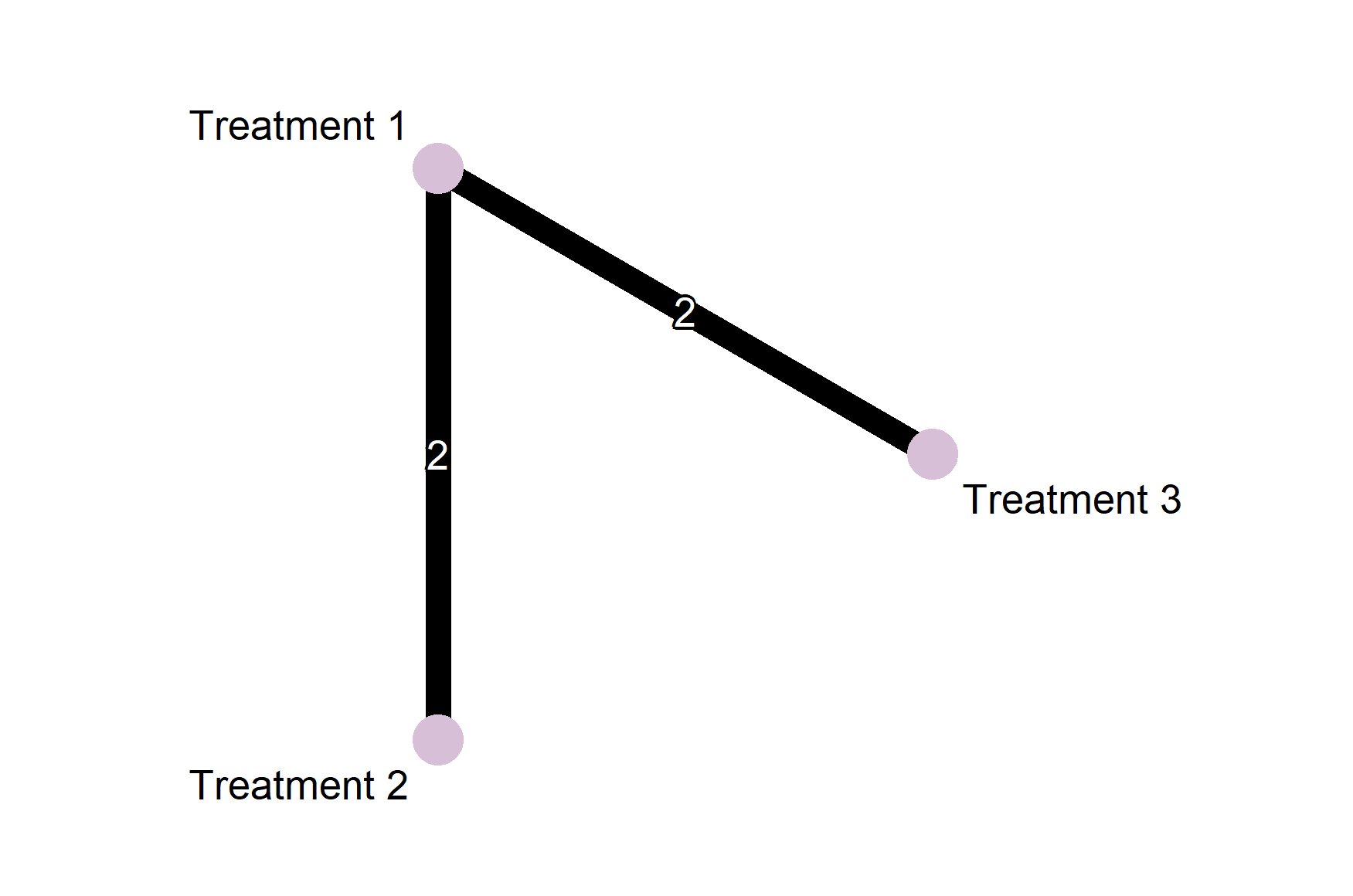}
    \caption{Network plot showing the network structure used in the simulation. Numbers on edges represent the number of studies making that direct comparison.}
    \label{fig:simnet}
\end{figure}

Since \citet{shen_two-stage_2025} found that network structure did not impact the performance of ITR NMA, we use a single network structure for our simulation. We simulate data from four studies making up the network connecting three treatments shown in Figure \ref{fig:simnet}, where each study has a total sample size of 500, since a small network is realistic in the context of ITR NMA. The studies labelled 1 and 2 compared treatments 1 and 2, and the studies labelled 3 and 4 compared treatments 1 and 3, that is, we have $\ti{1} = 1$ for all $i = 1, \dots, 4$, $\ti{2} = 2$ for $i = 1,2$ and $\ti{2} = 3$ for $i = 3,4$. For each data-generating mechanism, studies and corresponding individual-level data were simulated 5000 times. 

We considered a total of three DGMs that differ in the number of effect modifiers $Q$ and the presence of between-study heterogeneity. In particular, we study a simple scenario with $Q = 1$ under the presence and absence of between-study heterogeneity (DGMs A and B, respectively) and a scenario with no between-study heterogeneity where $Q=10$, which is similar to the motivating data setting (DGM C). For DGMs A and B, we simulate two covariates for each participant, $\Xijk{1}$ and $\Xijk{2}$. For DGM C, we simulate 10 covariates, $\Xijk{p}$ for $p = 1, \dots, 10$.  The covariates were then used to simulate treatment assignment, response, and missingness for each individual. The set of covariates involved in the true reference, blip, treatment assignment, and missingness mechanisms for each DGM are summarized in Table \ref{tab:mechanisms}. The distributions we used to simulate covariates and the details of each mechanism including parameter values are given in SI \S 3.

\begin{table}[tb]
    \centering
        \caption{Summary of data generating mechanisms (DGMs), including the covariates involved in each mechanism in the data generation process for all data generating mechanisms investigated in the simulation.}
    \label{tab:mechanisms}
    \begin{tabular}{c|ccccc}
        DGM & $\tau$ & Reference ($\xijb$) & Blip ($\xijd$) & Treatment assignment ($\xijt$) & Missingness ($\xijm$)  \\ \hline
        A &  $0.3$ & $\Xijk{1}, \Xijk{2}, \Xijk{1}^2$ &  $\Xijk{1}$ & $\Xijk{1}, \Xijk{2}$ & $\Xijk{1}, \aij$ \\
        B &$0$ & $\Xijk{1}, \Xijk{2}, \Xijk{1}^2$ &  $\Xijk{1}$ & $\Xijk{1}, \Xijk{2}$ & $\Xijk{1}, \aij$ \\
        C &  $0$ & $\Xijk{p}, p = 1, \dots, 10, \Xijk{1}^2$ & $\Xijk{p}, p = 1, \dots, 10$ & $\Xijk{1}, \Xijk{2}$ & $\Xijk{1}, \aij$
    \end{tabular}
\end{table}

\subsection{Methods}

We analysed each simulated dataset using a two-stage ITR NMA model as described in Section \ref{sec:nmamodel}. The combinations of investigated DGM, stage one estimation method, specification of the reference model, covariance structure for stage one estimates, and prior for $\tau$ are summarised in Table \ref{tab:simscenarios}. Q-learning with complete cases only and BBdWOLS with MAR outcome weights were used to obtain stage one estimates. Both approaches used correct specification of the blip models. In BBdWOLS, the weight models were always correctly specified. We varied the specification of the reference model for both Q-learning and BBdWOLS. Misspecification of the reference model was achieved by omitting $\xijo^2$ from the regression, which avoids violating the no unmeasured confounders assumption. We used $1999$ Bayesian Bootstrap iterations for DGMs A and B and 2999 for DGM C as it has many more parameters to be estimated and thus requires more iterations to converge. We used the posterior mean as the point estimate for $\ddeltai$.

We investigated the use of either the full estimated covariance matrix of $\hddeltai$ or the sparse version used in the model formulation of \citet{shen_two-stage_2025}. The sparse approach results in a diagonal covariance matrix in all DGMs because each study compared only two treatments. We also investigate the use of two different priors for $\tau$ in DGM A. We use a half-normal distribution family with different values of the scale parameter $\sigma_{\tau}$ \citep{rover_weakly_2021}. For the more vague prior, we set $\sigma_{\tau}$ to 0.51, which gives a prior median of 0.344 and 95\% percentile of 1. For the more informative prior, we set $\sigma_{\tau}$ to 0.45, which gives a prior median of 0.303 and 95\% percentile of 0.882. 

The NMA models were fit using Rstan \citep{stan_development_team_rstan_2024}. We used independent normal priors for the elements of $\ppsi$ with standard deviation set to 10 for all three DGMs. Each model was fit by running four independent MCMC chains with a total of $10,000$ iterations for DGM A with a full covariance specification and $20,000$ with a sparse specification, $8,000$ for DGM B, and $2,000$ iterations for DGM C. Convergence was assessed by ensuring the effective sample size for each parameter was at least 400 and R-hat was less than 1.01 \citep{vehtari_rank-normalization_2021, stan_development_team_rstan_2024}. The first half of the iterations from each chain were discarded as burn-in, and models which did not meet the convergence criteria were removed from the analysis. The number of iterations was chosen to ensure that a majority of models converged. The posterior mean was used as the point estimate and 95\% quantile-based CrIs were computed by taking the 0.025 and 0.975 quantiles of the posterior samples for all parameters of interest. 

\begin{table}[tb]
    \centering
        \caption{Summary of simulation scenarios.}
    \label{tab:simscenarios}
    \begin{tabular}{cllll}
        DGM & Stage one estimation & Reference model specification & $\Si$ & Prior for $\tau$ \\ \hline
        A & Q-learning with complete cases & Correct & Full & Vague \\
        A & Q-learning with complete cases & Incorrect & Full & Vague \\
        A & BBdWOLS, MAR outcome weights & Correct & Full & Vague \\ 
        A & BBdWOLS, MAR outcome weights & Correct & Full & Informative \\
        A & BBdWOLS, MAR outcome weights & Correct & Sparse & Vague \\
        A & BBdWOLS, MAR outcome weights & Correct & Sparse & Informative \\
        A & BBdWOLS, MAR outcome weights & Incorrect & Full & Vague \\ 
        A & BBdWOLS, MAR outcome weights & Incorrect & Full & Informative \\
        A & BBdWOLS, MAR outcome weights & Incorrect & Sparse & Vague \\
        A & BBdWOLS, MAR outcome weights & Incorrect & Sparse & Informative \\
        \hline
        B & BBdWOLS, MAR outcome weights & Correct & Full & N/A \\
        B & BBdWOLS, MAR outcome weights & Correct & Sparse & N/A \\\hline
        C & BBdWOLS, MAR outcome weights & Correct & Full & N/A \\
        C & BBdWOLS, MAR outcome weights & Correct & Sparse & N/A
    \end{tabular}
\end{table}

\subsection{Estimands and Performance Measures}

Primary interest is in the estimation of the meta-population blip parameters which define the estimated optimal ITR. We thus investigate the bias and empirical standard error (SE) of the elements of $\ppsi$ as well as the coverage probability of 95\% quantile-based CrIs, with formulas for each performance measure given in SI \S 2.2. In random-effects models, we are interested in the estimation of the between-study heterogeneity $\tau$, and investigate the bias and empirical SE in estimating this parameter. We also counted the number of non-converged models for each approach.

\subsection{Results: DGMs A and B (Low-Dimensional)}

The percent bias, empirical SE, and coverage probability for the meta-population blip parameters for all scenarios in DGM A are shown in Table \ref{tab:blip-sim-results}. First, we focus on comparing the use of Q-learning to BBdWOLS in stage one in DGM A with a full covariance matrix and $\sigma_\tau = 0.51$. As expected, Q-learning gives biased blip parameter estimates (absolute bias of 8.7 to 12.7\%) and 95\% CrIs have very low coverage probability for some parameters when the reference model is misspecified. BBdWOLS has bias values less than 1\% even when the reference model is misspecified, and has coverage probabilities of 95\% CrIs ranging from 0.96-0.97. The empirical SE of Q-learning is slightly lower than that of BBdWOLS for the main effect parameters when the reference model is correctly specified. 

Now, we focus on the impact of correctly specifying the reference model in BBdWOLS in DGM A. The bias in the blip parameters is slightly increased for settings when the reference model was misspecified, but the magnitude is still low (the largest value is $0.905\%$). Correctly specifying the reference model leads to lower empirical SEs. There is not a clear impact on the coverage probability of correct reference model specification. 

The impact of the structure of the covariance matrix of $\hddeltai$ seems to depend whether between-study heterogeneity is present or not. In DGM A, there is no clear pattern of the covariance structure used on bias, empirical SE, or coverage probability. However, in DGM B where there is no between-study heterogeneity, the sparse covariance structure leads to slightly worse bias for most parameters and higher empirical SE compared to the full covariance structure. In DGM B, regardless of the covariance structure used, the bias is very low, less than 0.05\%. In DGM A, between 195 and 670 out of the 5000 simulation repetitions that used the sparse covariance matrix did not converge, while only 65 or fewer repetitions did not converge for the settings with the full covariance matrix. In DGMs B and C, all models converged in all simulation iterations. Overall, we see that in low-dimensional settings, using the full covariance structure may reduce bias and improve efficiency when there is no between-study heterogeneity present, and can improve convergence in the presence of between-study heterogeneity. 

DGM A is the only mechanism with between-study heterogeneity. The prior distribution used for the between-study heterogeneity parameter in DGM A does not have a clear impact on bias or empirical SE of the blip parameters. Settings that used the more vague prior ($\sigma_\tau = 0.51)$ had higher over-coverage of blip parameters compared to the more informative prior. In general, all approaches had upwards bias in the estimation of $\tau$, ranging from 14.2-33.4\%. Misspecifying the reference model increases the bias and empirical SE of this parameter. Using the larger prior standard deviation also lead to higher bias and empirical SE. Using the sparse variance-covariance matrix lead to higher empirical SE and did not have a clear impact on the bias. 

\begin{sidewaystable}[ht]
\centering
\caption{Bias, empirical standard error (SE) and coverage probability (CP) of 95\% credible intervals (CrI) under a correct or incorrectly specified reference model across different methods and modelling choices for the blip parameters $\psi_{21,0}, \psi_{21,1}, \psi_{31,0}$ and $\psi_{31,1}$ in DGMs A, B, and C. Simulation results for the remaining blip parameters for DGM C are described in supplementary information.} 
\label{tab:blip-sim-results}
\begin{tabular}{llllr|rrrr|rrrr|rrrr}
  \hline
DGM & Stage one & Reference model & $\Si$ & $\sigma_\tau$ & \multicolumn{4}{c|}{\% Bias} & \multicolumn{4}{|c|}{Empirical SE} & \multicolumn{4}{|c}{CP of 95\% CrI} \\ 
& & & & & $\psi_{21,0}$ & $\psi_{31,0}$ & $\psi_{21,1}$  & $\psi_{31,1}$ & $\psi_{21,0}$ & $\psi_{31,0}$ & $\psi_{21,1}$  & $\psi_{31,1}$& $\psi_{21,0}$ & $\psi_{31,0}$ & $\psi_{21,1}$  & $\psi_{31,1}$\\
  \hline
  A & BBdWOLS & Correct & Full & 0.51 & -0.14 & 0.09 & -0.30 & 0.18 & 0.289 & 0.290 & 0.214 & 0.219 & 0.97 & 0.97 & 0.97 & 0.96 \\ 
  A & BBdWOLS & Not correct & Full & 0.51 & 0.38 & 0.61 & -0.75 & -0.42 & 0.389 & 0.390 & 0.220 & 0.225 & 0.96 & 0.96 & 0.97 & 0.97 \\ 
  A & Q-learning& Correct & Full & 0.51 & -0.16 & 0.11 & -0.27 & 0.17 & 0.286 & 0.286 & 0.214 & 0.218 & 0.97 & 0.97 & 0.97 & 0.96 \\ 
  A & Q-learning& Not correct & Full & 0.51 & 12.67 & 8.70 & -10.45 & -9.94 & 0.409 & 0.411 & 0.222 & 0.226 & 0.74 & 0.73 & 0.96 & 0.96 \\ 
  A & BBdWOLS& Correct & Full & 0.45 & -0.13 & 0.08 & -0.29 & 0.15 & 0.289 & 0.291 & 0.214 & 0.219 & 0.97 & 0.97 & 0.96 & 0.96 \\ 
  A & BBdWOLS &Not correct & Full & 0.45 & 0.39 & 0.60 & -0.73 & -0.44 & 0.389 & 0.390 & 0.220 & 0.225 & 0.96 & 0.95 & 0.97 & 0.96 \\ 
  A & BBdWOLS &Correct & Sparse & 0.51 & -0.14 & 0.08 & -0.29 & 0.27 & 0.289 & 0.291 & 0.214 & 0.219 & 0.97 & 0.97 & 0.97 & 0.96 \\ 
  A & BBdWOLS &Not correct & Sparse & 0.51 & 0.39 & 0.62 & -0.91 & -0.26 & 0.389 & 0.391 & 0.220 & 0.225 & 0.96 & 0.96 & 0.98 & 0.97 \\ 
  A & BBdWOLS &Correct & Sparse & 0.45 & -0.14 & 0.09 & -0.38 & 0.27 & 0.289 & 0.291 & 0.214 & 0.219 & 0.97 & 0.97 & 0.96 & 0.96 \\ 
  A & BBdWOLS &Not correct & Sparse & 0.45 & 0.40 & 0.60 & -0.70 & -0.37 & 0.388 & 0.390 & 0.220 & 0.226 & 0.96 & 0.95 & 0.97 & 0.96 \\ 
   \hline
   B & BBdWOLS& Correct & Full &  & 0.07 & -0.05 & -0.05 & 0.05 & 0.198 & 0.198 & 0.039 & 0.038 & 0.94 & 0.94 & 0.94 & 0.94 \\ 
  B & BBdWOLS&Correct & Sparse &  & 0.07 & -0.05 & -0.07 & 0.09 & 0.201 & 0.200 & 0.039 & 0.039 & 0.94 & 0.94 & 0.94 & 0.94 \\ \hline
  C & BBdWOLS & Correct & Full & &-0.81 & 2.20 & -0.73 & -0.49 & 1.272 & 1.271 & 0.209 & 0.207 & 0.92 & 0.92 & 0.90 & 0.91 \\ 
  C & BBdWOLS & Correct & Sparse & & -0.81 & 2.26 & -0.56 & -0.19 & 1.285 & 1.293 & 0.226 & 0.224 & 0.93 & 0.93 & 0.92 & 0.92 \\ \hline
\end{tabular}
\end{sidewaystable}

\subsection{Results: DGM C (Moderate-Dimensional)} 

DGM C has significantly more blip parameters to consider compared to DGMs A and B. We display results for two out of the total 22 blip parameters in Table \ref{tab:blip-sim-results}. Additional results are described in SI \S 3. The bias is not high regardless of the covariance matrix used, with most parameters having bias below 1\%, and the maximum relative bias of 3.59\% occurring for the parameter $\psi_{21,9}$ when the full covariance matrix was used. The coverage probability of CrIs is below 0.95 for all parameters regardless of the covariance matrix used. The sparse specification tends to give coverage closer to the nominal level although this is not seen for all parameters. The empirical SE of all parameters is higher for the sparse covariance matrix compared to the full covariance matrix, showing that the full covariance matrix specification can increase efficiency when there are many effect modifiers.

\begin{table}[tb]
\centering
\caption{Simulation results for the common between-study heterogeneity parameter $\tau$.}
\label{tab:tau-sim-res}
\begin{tabular}{llllrrr}
  \hline
DGM & Stage one & Reference model & $\Si$ & $\sigma_\tau$ & \% Bias (SE) & Empirical SE (SE) \\ 
  \hline
A & BBdWOLS & Correct & Full & 0.45 & 14.2 (0.5) & 0.113 (0.001)\\ 
  A & BBdWOLS & Correct & Full & 0.51 & 18.7 (0.6) & 0.119 (0.001) \\ 
  A & BBdWOLS & Correct & Sparse & 0.45 & 14.5 (0.6) & 0.118 (0.001) \\ 
  A & BBdWOLS & Correct & Sparse & 0.51 & 19.2 (0.6) & 0.124 (0.001)\\ 
  A & BBdWOLS & Not correct & Full & 0.45 & 17.1 (0.5) & 0.115 (0.001)\\ 
  A & BBdWOLS & Not correct & Full & 0.51 & 22.4 (0.6) & 0.122 (0.001)\\ 
  A & BBdWOLS & Not correct & Sparse & 0.45 & 16.0 (0.6) & 0.126 (0.001)\\ 
  A & BBdWOLS & Not correct & Sparse & 0.51 & 21.5 (0.6) & 0.134 (0.001)\\ 
  A & Q-learning & Correct & Full & 0.51 & 18.0 (0.6) & 0.118 (0.001)\\ 
  A & Q-learning & Not correct & Full & 0.51 & 33.4 (0.7) & 0.142 (0.001)\\ 
   \hline
\end{tabular}
\end{table}

\section{Data Analysis}\label{sec:dataresults}

\subsection{Methods}

We applied the fully Bayesian, doubly-robust ITR NMA method to the motivating dataset to estimate an optimal ITR for MDD. We use the negative HRSD-17 score as the response, so a larger value indicates less severe depression and a more preferred treatment. Since each of the three studies included a SER treatment arm, we use this as the reference treatment (treatment 1) at both the study and meta-population level. The demographic and clinical variables that were available in all studies that are expected to predict the outcome and/or act as effect modifiers are race, age, sex, marital status, years of formal education completed, employment status, household size, age of MDD onset, number of depression episodes, whether the current episode is chronic, and the baseline HRSD-17 score \citep{kessler_using_2017, perlman_systematic_2019}. All of these variables are included in the reference model. In the blip model, we include all of these variables except for race and chronicity of current episode because it is not recommended to tailor treatment based on race, and in the REVAMP study, only 11 patients had their current episode categorized as chronic, making it difficult to estimate treatment interactions with this variable. Details about the coding of each variable are given in SI \S 4.

There are both missing outcomes and missing covariates in the data. We use simple imputation to deal with missing covariates in this analysis, corresponding to the assumption that missing covariate values are not related to the outcome or to any other covariates. We imputed the mean for continuous covariates and the most commonly observed value for categorical and binary variables.

We use MAR outcome weights in BBdWOLS as defined in Section \ref{sec:BBdWOLS-miss}. We reflect the unique design of each study described in Section \ref{sec:data} as well as pragmatic considerations in the specification of the treatment assignment and missingness models used to determine the weights. The specific models for each study are given in SI \S 4. EMBARC and REVAMP exhibited extreme weights for some individuals in some bootstrap iterations, so we trimmed weights $w_{ijl}$ to the 99th percentile in these studies. 

We obtained study-level ITR estimates using BBdWOLS with MAR outcome weights and used the full specification of $\Si$ in the ITR NMA model as described in Section \ref{sec:nmamodel}. We used a common-effects NMA model since most treatment comparisons were only observed once, making it difficult to estimate between-study variability without overly relying on prior information. We used independent normal priors for the elements of $\ppsi$ with standard deviation set to 13.25, as this was deemed sufficiently vague while still incorporating information about the possible range of the responses.

First stage estimates and corresponding covariance matrices were obtained using the BBdWOLS R function available at \href{https://github.com/augustinewigle/bbdwols}{github.com/augustinewigle/bbdwols} and the fwb R package \citep{greifer_fwb_2025}.  The NMA model was fit using Stan via the RStan package \citep{stan_development_team_rstan_2024, stan_development_team_stan_2025} with four independent chains with 1000 iterations preceded by 1000 warm-up iterations. Convergence was assessed by ensuring a high effective sample size and an R-hat below 1.01 for each parameter \citep{vehtari_rank-normalization_2021, stan_development_team_stan_2025}. We also performed several sensitivity analyses using different weighting approaches in BBdWOLS and sparse covariance matrices in the NMA model. The results of the sensitivity analyses are shown in SI \S 4.

\subsection{Results}

\begin{figure}[tb]
    \centering
    \includegraphics[width=0.75\linewidth]{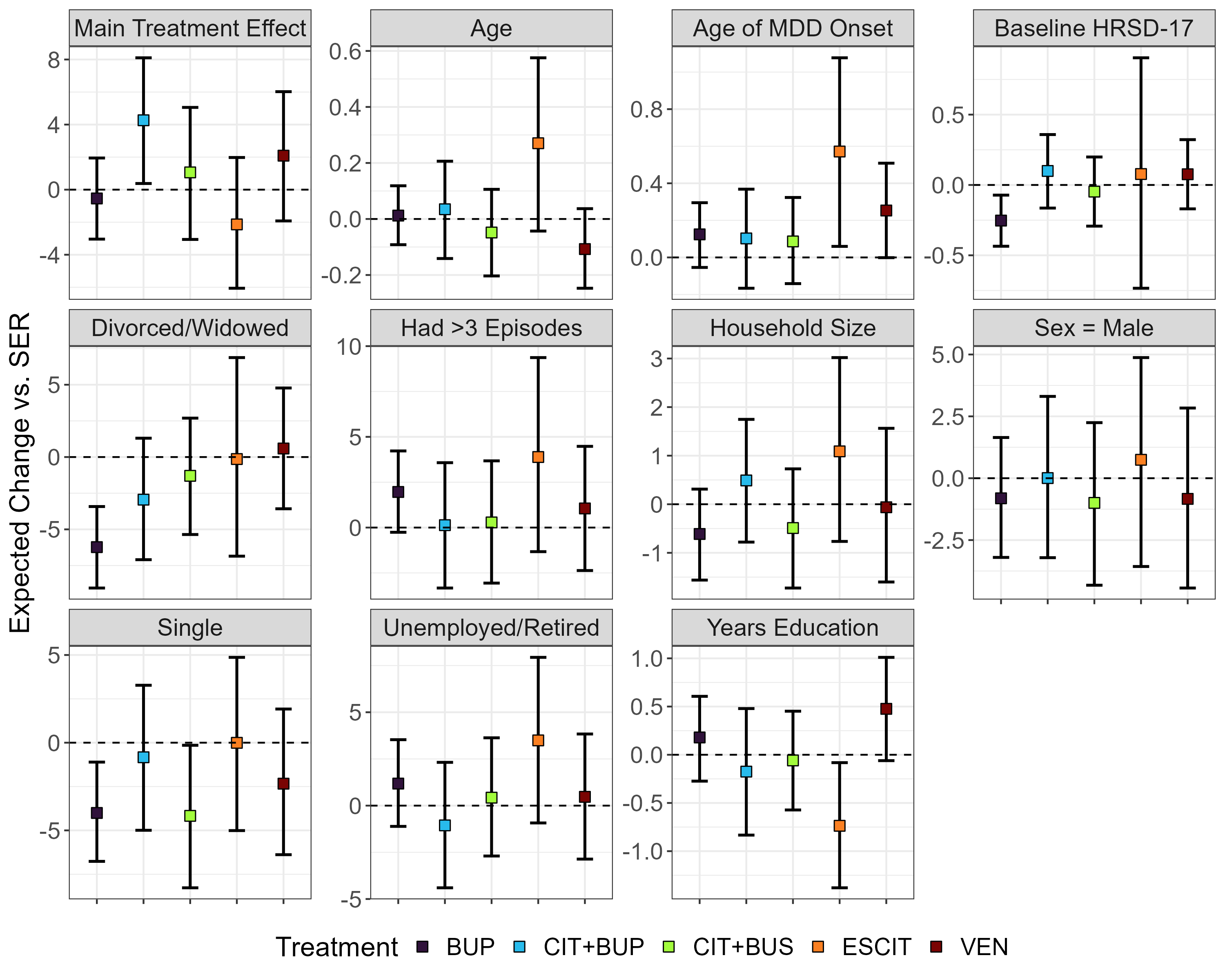}
    \caption{Blip parameter estimates for MDD using BBdWOLS with MAR outcome weights and a full covariance matrix specification.}
    \label{fig:goodres}
\end{figure}


The point estimates and 95\% CrIs for the meta-population blip parameters describing the optimal ITR for MDD are shown in Figure \ref{fig:goodres}. Out of all the treatments, the coefficients associated with BUP are the most precisely estimated while ESCIT has the least precisely estimated coefficients. This can be understood as BUP was investigated in all three studies, with a total of 246 patients receiving this treatment, while ESCIT was only received by 53 patients in one study. Many coefficients' 95\% CrIs cover zero, with seven exceptions (main effect of CIT+BUP, three effect modifiers' interactions with BUP, one effect modifier interaction with CIT+BUS, two effect modifiers' interactions with ESCIT).

It is important to note that the CrIs displayed in Figure \ref{fig:goodres} reflect comparisons to the treatment SER, and it is not possible to determine from this plot whether any other treatment comparisons meaningfully differ from one another. For example, although the CrIs for $\psi_{\text{Escit Ser, Age}}$ and $\psi_{\text{Ven Ser, Age}}$ 
overlap in Figure \ref{fig:goodres}, computing the posterior of the interaction of age with ESCIT versus VEN according to the consistency equation, that is, for each MCMC iteration, computing
\begin{equation*}
    \psi_{\text{Escit Ven}, \text{Age}} = \psi_{\text{Escit Ser}, \text{Age}} - \psi_{\text{Ven Ser}, \text{Age}},
\end{equation*}
leads to a 95\% CrI that does not contain zero.

Each coefficient in the blip expresses the interaction of one covariate at a time. A summary of how these interactions impact treatment performance for a specific covariate profile may be more useful for individual decision-making than looking at the blip parameter estimates. To support this aim, we can calculate the expected relative effect for each treatment compared to SER given covariates $\bm{x}$ at each MCMC iteration according to
\begin{equation*}
    \ppsi_g'\bm{x}, \quad g \in (\text{BUP}, \text{CIT+BUP}, \text{CIT+BUS}, \text{ESC}, \text{VEN}),
\end{equation*}
and examine the posterior distributions of the expected relative effects. 

\begin{figure}[tb]
    \centering
    \includegraphics[width=0.5\linewidth]{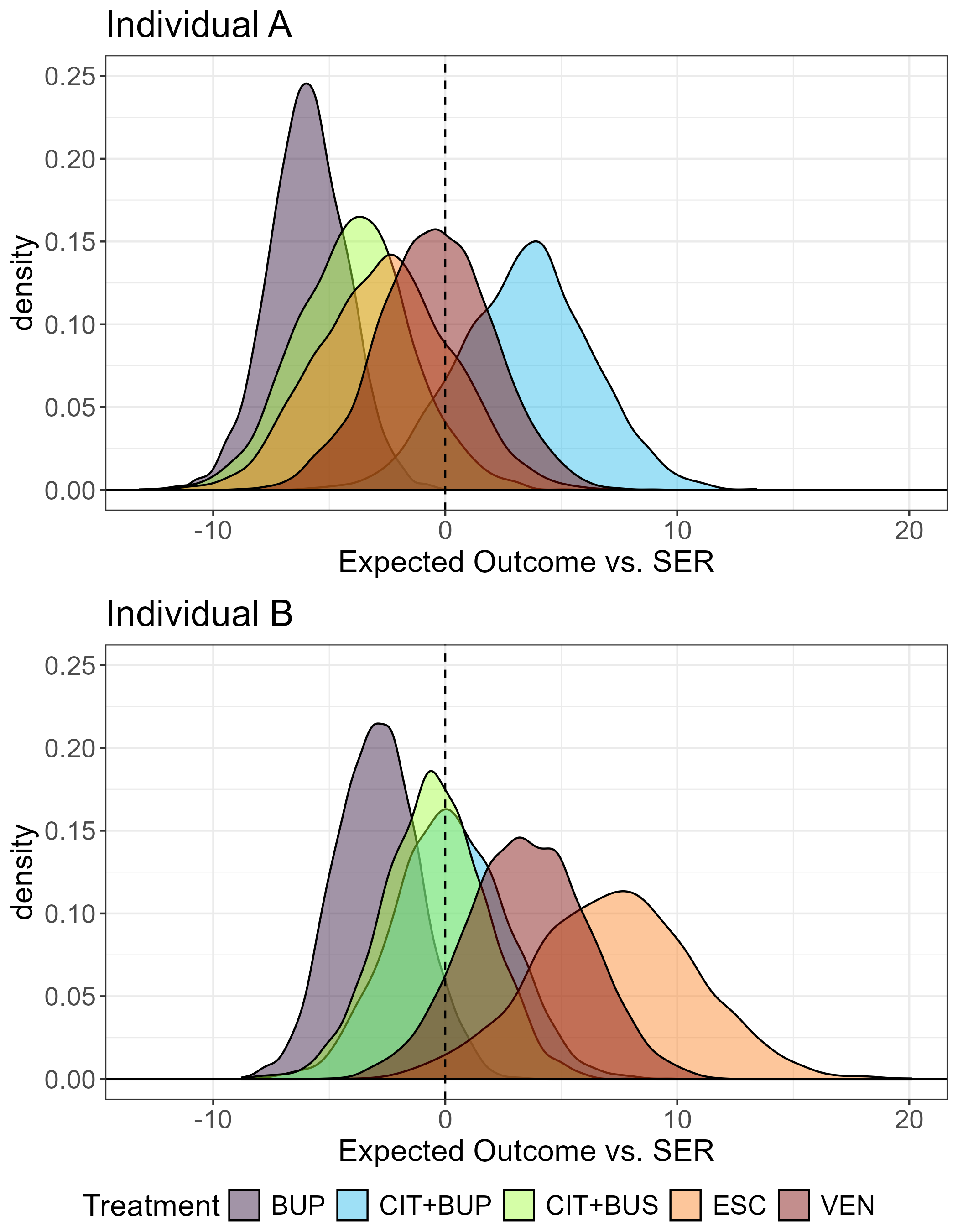}
    \caption{Posterior distribution of expected relative effects for individuals A and B with different covariate profiles.}
    \label{fig:indlposts}
\end{figure}

To illustrate, we describe two hypothetical individuals, A and B, with different covariate profiles. The individuals' covariate profiles are shown in SI \S 4. In summary, the individuals differ in their age, baseline HRSD-17, years of education, employment status, marital status, number of MDD episodes, and age at onset of MDD. The posterior distributions of the relative effects for each individual are shown in Figure \ref{fig:indlposts}. It is clear that the treatments have different relative performance between the individuals, for example, for Individual A, the treatment with the largest posterior mean is CIT + BUP, while for Individual B, it is ESCIT. Examining the posterior distributions rather than just the expected relative effects has the advantage of conveying uncertainty through the width and overlap of the distributions, which is particularly useful when considerations beside outcome magnitude are important, such as cost. 

\section{Discussion} \label{sec:disc}

ITR NMA is a powerful tool for estimating optimal ITRs from a wide set of competing treatments. In this work, we introduced BBdWOLS, a doubly-robust Bayesian approach to estimating study-level ITRs that accommodates MAR outcomes. We proposed the use of the full variance-covariance matrix of study-level ITR estimates in the two-stage ITR NMA model. In a simulation study, we showed that using BBdWOLS in stage one of the ITR NMA is robust and that using the full variance-covariance matrix in stage two can reduce the bias and empirical SE of blip parameter estimators and improve convergence. We applied our method to a network of three studies to estimate an optimal ITR for MDD.

One-stage and two-stage approaches to ITR NMA are possible \citep{shen_two-stage_2025}. A benefit of one-stage approaches is that the exact likelihood of the data can be used, while two-stage approaches rely on a normality assumption \citep{riley_individual_2020}. Differences between one and two-stage approaches are typically minimal except in small within-study sample sizes or with rare events, and it is easier to make modelling mistakes when attempting to specify a one-stage model \citep{riley_individual_2020, hua_one-stage_2017}. In the two-stage approach, study-specific ITR estimates can be published without revealing IPD, making it possible for readers to reproduce the ITR NMA results. This can be seen as an advantage over one-stage approaches, where IPD would need to be released to reproduce any part of the results. 

The regression models fit in stage one of the proposed ITR NMA approach must be of the same form in all studies. This can present challenges if important prognostic or effect-modifying variables are not reported in all studies, or if only one value of an important covariate is observed in a study, preventing estimation of its interaction with treatments. The latter challenge has been explored in the context of ITR meta-analysis through careful model parameterization, and could be further extrapolated to the case of NMA \citep{shen_sparse_2025}. Adequate reporting of important variables in studies is crucial to enable the correct specification of ITR models and synthesis with ITR NMA.

We extended dWOLS to account for MAR outcomes in this work, as missing outcomes are more common than missing covariates in the context of randomized study data. However, missingness in covariates is also a common problem, even in randomized studies. We performed a sensitivity analysis in the MDD data where we modelled the missingness of outcome or covariates as a function of completely observed covariates. We found that this model did not perform as well as the presented model, possibly because the treatment assignment and missingness models could not be correctly specified when we were restricted to completely observed covariates. Further development of these methods to account for missing outcomes and covariates simultaneously will enable more robust applications and conclusions.

In Section \ref{sec:dataresults}, we illustrated how a set of CrIs for main effects and interactions relative to one treatment may not paint a full picture of the ITR. Given that there may be tens or even hundreds of parameters characterizing an ITR, methods to summarize the results in a useful and statistically sound way are needed. We have suggested computing the posterior distribution of the expected relative effects for each treatment for a given covariate profile. In random-effects models, computing the posterior predictive distribution of the relative effects may more accurately characterize the uncertainty in the expected treatment effects for a new patient \citep{rosenberger_predictive_2021}. Ranking metrics, which produce an ordered list of competing treatments based on preference, are often used to summarise the results of a standard NMA \citep{salanti_introducing_2022}. Extensions of ranking metrics that can be applied to ITR NMA may further support decision-making \citep{wigle_personalized_2026}. 

A challenge in ITR NMA is finding appropriate studies. Access to relevant study data at the individual level is often limited due to privacy concerns. Even when IPD are available, studies may report outcomes on different scales, for example, depression severity is reported using the HRSD-17 in the motivating data, but other scales for depression severity are also commonly reported, such as the self-reported or clinician-administered Quick Inventory of Depressive Symptomatology \citep{furukawa_assessment_2010}. Relevant studies may compare non-overlapping sets of treatments, preventing simultaneous synthesis. A limited sample size in terms of the studies that can be pooled in ITR NMA can lead to sensitivity to prior specification and difficulty in estimating between-study heterogeneity, as we observed in the simulation study and data analysis. Some literature has been dedicated to prior specification of the between-study heterogeneity parameter in meta-analysis (e.g., \citet{turner_predicting_2012, rover_weakly_2021}) as well as in NMA under the homogeneous variance assumption \citep{van_valkenhoef_automating_2012}. Further investigation into prior specification of $\SSigma$, including under different covariance structures and in the specific context of ITR NMA, is warranted. Other challenges stemming from small sample size include reduced power and a limited set of treatments to compare. To increase the number of studies that can be pooled in an ITR NMA, methodological developments that address challenges related to outcome scales and non-overlapping treatment sets are needed.

The proposed ITR NMA approach extends existing NMA methods in important ways. First, the majority of previous work in NMA has focused on synthesizing evidence from randomized studies to avoid bias due to confounding. Existing work on incorporating non-randomized studies into NMA either do not estimate causal effects \citep{ades_twenty_2024} or attempt to model the bias in non-randomized studies \citep{hamza_synthesizing_2023, hussein_hierarchical_2023}. Our approach provides a coherent framework to estimate causal effects from both randomized and non-randomized studies in NMA. Further, it is doubly-robust, protecting against model misspecification, which is particularly important when non-randomized studies for which the treatment assignment mechanism is not known are included. Second, we have shown how our approach can be used to estimate conditional relative treatment effects that can support individual decision making. Conditional relative treatment effects can be obtained from other IPD NMA approaches and the reporting of relative effects and treatment rankings at different covariate levels has been encouraged \citep{donegan_combining_2013-1, freeman_framework_2018, riley_individual_2020, riley_using_2023}, but this practice has not yet been widely adopted. We hope that our approach can further illuminate how NMA can enable personalized decision-making.

\section{Funding}

AW is supported by NSERC (grant number PDF - 598932 - 2025).
EEMM is a CIHR Canada Research Chair (Tier 1) in Statistical Methods for Precision Medicine.

\bibliographystyle{plainnat}
\bibliography{awigle-freiburg, awigle-postdoc}

\end{document}